# Origin of the large positive magnetoresistance
# in Ge$_{1-x}$Mn$_x$ granular thin films

Yuki K. Wakabayashi,[1] Ryota Akiyama,[2] Yukiharu Takeda,[3] Masafumi Horio,[2]
Goro Shibata,[2] Shoya Sakamoto,[2] Yoshisuke Ban,[1] Yuji Saitoh,[3] Hiroshi Yamagami,[3,4]
Atsushi Fujimori,[2] Masaaki Tanaka,[1,5] and Shinobu Ohya[1,5]

[1]*Department of Electrical Engineering and Information Systems,*
*The University of Tokyo, 7-3-1 Hongo, Bunkyo-ku, Tokyo 113-8656, Japan*
[2]*Department of Physics, The University of Tokyo, Bunkyo-ku, Tokyo 113-0033, Japan*
[3]*Materials Sciences Research Center, JAEA, Sayo, Hyogo 679-5148, Japan*
[4]*Department of Physics, Kyoto Sangyo University, Motoyama, Kamigamo, Kita-Ku,*
*Kyoto 603-8555, Japan*
[5]*Center for Spintronics Research Network, Graduate School of Engineering,*
*The University of Tokyo, 7-3-1 Hongo, Bunkyo-ku, Tokyo 113-8656, Japan*

Abstract

Ge$_{1-x}$Mn$_x$ (GeMn) granular thin films are a unique and promising material for spintronics
applications due to large positive magnetoresistance (MR). Previous studies on GeMn
have suggested that the large MR is related to nanospinodal decomposition of GeMn into
Mn-rich ferromagnetic nanoparticles and Mn-poor paramagnetic matrix. However, its
microscopic origin of the MR has not been clarified yet. Here, using X-ray magnetic
circular dichroism (XMCD), which is extremely sensitive to the local magnetic state of
each atom, we investigate the magnetic properties of the nanoparticles and the matrix in
GeMn separately. We find that the MR ratio is proportional to the product of the
magnetizations originating from the nanoparticles and the matrix. This result indicates
that spin-polarized holes in the nanoparticles penetrate into the matrix and that these holes
undergo spin-disorder magnetic scattering by the paramagnetic Mn atoms in the matrix,
which induces the large MR.

Ge$_{1-x}$Mn$_x$ (GeMn) granular thin films have attracted much interest for future spintronic applications due to their large positive magnetoresistance (MR) up to ~280% (under 5 T at 40 K) and due to the compatibility with existing semiconductor technology [ 1 – 8 ]. Generally, MR observed in granular materials is explained by giant magnetoresistance (GMR) [9] or magnetic-field-dependent avalanche breakdown [10,11]. However, in GeMn, the sharp enhancement of MR at very low temperature and its peculiar spike-shaped magnetic-field dependence cannot be explained by those conventional effects. Previous studies on GeMn have suggested that the large MR is related to the nanoscale spinodal decomposition of GeMn into ferromagnetic (FM) metallic Mn-rich nanoparticles and paramagnetic (PM) Mn-poor matrix (Fig.1(a)) [1,2,6,7]. However, the microscopic origin of the MR has not been clarified yet. The large MR is only observed below the percolation temperature ($T_p \approx 10$ K) [7,12], below which spin-polarized holes in the nanoparticles penetrate into the matrix and are interconnected [1,2,6,7]. Thus, one can presume that the spin-polarized holes, which are extended from the nanoparticles into the matrix, are subject to spin disorder magnetic scattering by the PM Mn atoms in the Mn-poor matrix and induce the MR below $T_p$ [Figs. 1(b) and 1(c)]. Because this spin-dependent scattering occurs near the interfaces between the nanoparticles and the matrix, separate detection of the magnetic properties of the FM nanoparticles and the PM matrix near the interfaces is necessary to test this hypothesis. However, such a separate detection is difficult with conventional magnetization measurements using a superconducting quantum interference device (SQUID).

Soft X-ray magnetic circular dichroism (XMCD) is sensitive to the local magnetic state of each atom in magnetic films [13-18]. One can distinguish between different local magnetic states by the difference in the energy spectrum as well as the difference in the magnetic-field dependence of XMCD signals from each atom. Thus, by carefully analyzing the magnetic-field dependence of XMCD signals using various incident-photon energies, one can distinguish magnetic signals originating from the FM nanoparticles and the PM matrix. Another advantage of XMCD, especially in our study, is its probing depth. In the total electron yield (TEY) mode used for the present XMCD measurements, we detect signals originating from atoms located within 2–3 nm from the film surface [19]. As shown in Fig. 1(a), the nanoparticles are located at around 3-5 nm from the film surface of GeMn. Thus, one can selectively study the magnetic properties near the top interfaces of those nanoparticles. Because the scattering of charge carriers takes place near those interfaces, XMCD measurements are suitable for the investigation of spin-disorder magnetic scattering in GeMn.

In this Letter, we investigate the magnetic properties of the nanoparticles and the matrix separately and clarify the relationship between the MR and $T_p$, utilizing the advantages of XMCD and carefully analyzing the XMCD data. We used the twin-helical undulator beamline BL23SU of SPring-8, which enables us to perform efficient measurements of XMCD with various incident photon energies and magnetic fields at various temperatures [20]. We find that the MR ratio is proportional to the product of the magnetizations originating from the FM nanoparticles and the PM matrix below $T_p$. This indicates that the spin-polarized holes, which penetrate from the nanoparticles into the matrix, undergo spin-disorder magnetic scattering by the PM Mn atoms in the matrix, causing the large MR.

We grew 13-nm-thick epitaxial $Ge_{1-x}Mn_x$ granular thin films with the total Mn concentration $x$ of 0.09 and 0.14 on $p$ (or $n$)-type Ge(111) substrates using low temperature molecular-beam epitaxy (see Section I of Supplemental Material (SM) [21]). To avoid parallel conduction through the substrate, the samples for the magneto-transport measurements were grown on $n$-type substrates. This is because GeMn is $p$-type and the $p$-$n$ junction of $p$-GeMn/ $n$-Ge prevents the carrier diffusion from the GeMn layer to the substrate [22]. For other samples, after the growth of the $Ge_{1-x}Mn_x$ layer, we grew a 1.5-nm-thick Ge capping layer to prevent surface oxidation of the $Ge_{1-x}Mn_x$ layer. Our transmission microscope analyses showed that the GeMn films had sphere-like Mn-rich nanoparticles and a Mn-poor surrounding matrix (Fig. 1(a)). The nanoparticles are located at 3–5 nm from the film surface. They form in an amorphous metallic GeMn phase including $Mn_5Ge_3$ precipitates [2,7,23]. Meanwhile, the Mn-poor surrounding matrix has a diamond-type crystal structure (see Section II of SM [21]). Using spatially resolved energy dispersive X-ray spectroscopy, the local Mn concentration in the nanoparticles is roughly estimated to be ~60%, while the one in the matrix is to be ~6%.

Most of the 1.5-nm-thick Ge capping layer is naturally oxidized in the atmosphere. To remove this layer, the samples were briefly etched in a dilute HF solution prior to loading them into the XMCD vacuum chamber. Before the measurements, we carefully checked the absence of the oxide contamination at the sample surfaces using X-ray absorption spectroscopy (XAS) [24].

Figure 2 shows the Mn $L_{2,3}$-edge XAS $[(\mu^+ + \mu^-)/2]$ spectrum (a) and the XMCD (= $\mu^+ - \mu^-$) spectra (b) for the $Ge_{0.86}Mn_{0.14}$ film at 6 K with various magnetic fields applied perpendicular to the film surface. Here, $\mu^+$ and $\mu^-$ refer to the absorption coefficients for photon helicity parallel and antiparallel to the Mn $3d$ majority spin direction, respectively. In both XAS and XMCD spectra, one can see five peaks at the Mn $L_3$-edge (whose energies are referred to as $a$-$e$) [see also the insets in Figs. 2(a) and 2(b)] and two peaks at the Mn $L_2$-edge (whose energies are referred to as $f$ and $g$). When the XMCD spectra are normalized at $c$, the spectral line shape is changed with varying $H$, and the peak at $c$ becomes more dominant as $H$ increases, as shown in the inset of Fig. 2(b). While the XMCD intensities at $a$ and $b$ tend to saturate for $\mu_0 H = 7$ T, the one at $c$ does not. This indicates that the peaks at $a$ and $b$ have a certain amount of an FM component while the peak at $c$ mainly originates from the PM Mn atoms. The same features were also observed in the Ge$_{0.91}$Mn$_{0.09}$ film (see Section III of SM [21]). These results indicate that the XMCD signals have both PM and FM components.

By carefully analyzing the XMCD-$H$ curves measured at various energies and temperatures $T$ for both samples with $x = 0.09$ and 0.14, we separated the XMCD signals into the FM (-like) component, which saturates at high magnetic fields (> 6 T), and the PM component, which is linear in the range of $\mu_0 H$ from $-1$ to 1 T and follows the Langevin function (for detailed procedure, see Sections IV and V of SM [21]). One of the examples of the separation of the XMCD signal for Ge$_{0.86}$Mn$_{0.14}$ is shown in Fig. 3. At all temperatures, we succeeded in decomposing the XMCD-$H$ curves into the FM and PM components. Furthermore, all the XMCD-$H$ curves measured with different energies are expressed by the linear combination of these FM and PM XMCD-$H$ curves shown in Figs. 3(b) and 3(c) (see Fig. S5 in Section IV of SM [21]). This means that there exist only the two components (FM and PM components) in the XMCD signals. Moreover, the following results verify our decomposition procedure of the XMCD signals. The Curie plot ($H$/XMCD - $T$) of the derived PM component of the XMCD was linear, which is a typical PM behavior and confirms that the PM component is derived correctly in our study (see Section VI of SM [21]). In Fig. 3(b), the FM component of the XMCD becomes zero at 200 K, which means that the local ferromagnetism appears below 200 K [1,2,6,7]. This result is consistent with the previous studies on GeMn granular films [1,2,6,7]. Similar features were also observed in Ge$_{0.91}$Mn$_{0.09}$ (see Fig. S8 in Section V of SM [21]).

From the above analysis, we derived the FM and PM components of the XMCD signal at various energies for both samples, as shown by green and blue points in Fig. 4, respectively for both samples. The FM component of the XMCD spectra has a broad single negative peak at the Mn-$L_3$ edge. This is a typical feature that can be seen for the delocalized 3$d$ electrons of the FM Mn atoms in metallic materials. This confirms that the FM component indeed originates from the Mn-rich nanoparticles, each of which is locally metallic. Meanwhile, the PM component of the XMCD signal is attributed to the Mn-poor matrix. In fact, the derived XAS spectra of the PM Mn atoms have three peaks at the Mn-$L_3$ edge (see Section VII of SM [21]), which is consistent with the result of the first-principles calculation of the XAS spectrum of the Mn atoms that substitute the Ge sites in Ge$_{1-x}$Mn$_x$ [25]. The PM component of the XMCD spectra has three peaks at $c$, $d$, and $e$ at the Mn-$L_3$ edge (Fig. 4), which is a characteristic feature of the localized $3d$ state of the Mn$^{2+}$ ions with the magnetic moment of $5\mu_B$ [17,26,27]. From the Langevin function that expresses the PM component of the XMCD-$H$ curve at 6 K, the magnetic moment of the PM component is also estimated to be $5\mu_B$ [see Fig. S4(d) in Section IV of SM [21]]. These are characteristic features of the Mn atoms in insulating materials and are consistent with the insulating behavior of the matrix region of GeMn, which is evidenced by the variable range hopping transport observed in GeMn [2,7]. Also, the estimated spin and orbital magnetic moments of the Mn-rich nanoparticles and for the Mn-poor matrix support the above-mentioned assignment of the FM and PM components to the nanoparticles and matrix, respectively (see Section VIII of SM).

Because XMCD preferentially detects Mn atoms located near the top interfaces of the nanoparticles, the magnetic properties obtained by SQUID, which detects the magnetic properties of the entire film, are different from our XMCD results. While we do not see a hysteresis in the XMCD-$H$ curves (Fig. 3), it is observed in the SQUID measurements (see Section IX of SM [21]) [7,28]. Thus, in the nanoparticles, the magnetic properties at the interface are different from those at the core [29]. As mentioned below, holes located near these interfaces experience the spin-disorder magnetic scattering and thus have a key role for the large MR. This means that the XMCD measurements are uniquely suitable for the investigation of the spin-disorder magnetic scattering in GeMn.

Figures 5(a) and 5(b) show the MR ratio, defined as $[\rho(H) - \rho(0)]/\rho(0)$, for the Ge$_{0.86}$Mn$_{0.14}$ film (a) and Ge$_{0.91}$Mn$_{0.09}$ film (b). Here, $\rho(H)$ represents the resistivity of the GeMn films with $H$ applied perpendicular to the film surface. The magnetic-field dependence of the product of the FM and PM components of the XMCD intensity is also plotted. The MR curves for both samples show a spike-shaped curve at 6 K ($< T_p \approx$ 10 K), which is specific to the GeMn granular films. The MR shows a large enhancement below $T_p$. The MR ratio reaches 199% and 109% at 6 K ($< T_p$) when $\mu_0 H = 9$ T in Ge$_{0.86}$Mn$_{0.14}$ and Ge$_{0.91}$Mn$_{0.09}$, respectively. We see that the MR ratio is proportional to the product of the FM and PM components of the XMCD intensities (Fig. 5). This means that the MR is induced by the first order magnetic scattering of spin-polarized holes [30], which is expressed as

$$\text{MR} = -4P_{\text{FM}} \frac{M_{\text{FM}}(H)}{|M_{\text{FM}}(H)|} \frac{J_{\text{pd}}}{g\mu_B V} M_{\text{PM}}(H), \quad (1)$$

where $P_{\text{FM}}$, $J_{\text{pd}}$, $g$, $V$, $M_{\text{FM}}(H)$, and $M_{\text{PM}}(H)$ represent the spin polarization of holes in the FM regions, the $p$-$d$ exchange coupling constant between the holes and the PM Mn atoms, $g$-factor, field-independent part of the potential, magnetization of the FM Mn atoms, and magnetization of the PM Mn atoms. Generally, below $T_p$, spin polarized holes in the FM Mn-rich regions penetrate into the Mn-poor matrix and overlap with other holes that are extended from different Mn-rich nanoparticles [see Figs. 1(b) and 1(c)]. This induces the percolation and the long-range FM ordering [1,2,6,7,12]. Thus, our results indicate that the spin polarized holes, which are extended from the nanoparticles, undergo spin disorder magnetic scattering from the PM Mn atoms in the matrix below $T_p$ and that this scattering induces the large positive MR [1,2,6,7].

When $T > T_p$, the MR was significantly reduced and the shape of the MR curve was changed to concave (Fig. 5(a)). This means that the origin of the MR is different between $T > T_p$ and $T < T_p$. In GeMn, the resistivity has a bump at $T = T_p$ [2,7]. By increasing $H$, spins tend to be aligned and percolation can occur more easily, which leads to an increase in $T_p$. Thus, when $T > T_p$, the resistivity increases with increasing $H$, reflecting the approach of $T_p$ to the measurement temperature [1,2,7]. This induces the small positive MR when $T > T_p$.

In the case of (In,Mn)Sb, which is a similar granular system to GeMn, the spin polarization of holes is estimated to be $10 - 30$ % by the analysis of the MR curves using the above-mentioned magnetic scattering model [30]. Because the MR ratio of $Ge_{0.86}Mn_{0.14}$ is about ten times as large as that of (In,Mn)Sb, the spin polarization of holes in the nanoparticles is expected to be much larger in $Ge_{0.86}Mn_{0.14}$. By taking the same $J_{pd}/V$ value as (In,Mn)Sb (=0.17 [30]), the spin polarization of holes is estimated to be 64% from Eq. (1). This large spin-polarization is thought to be the origin of the large MR in GeMn below $T_p$, making GeMn a promising material for future spintronic applications. By the careful analysis of XMCD, we obtained the detailed magnetic properties of the Mn-rich nanoparticles and Mn-poor matrix separately. This unique method will be useful also for other granular material systems and will help to understand the mechanism of the MR and give us a clue to design how to increase the MR ratio in granular films.

In summary, we have carried out XMCD measurements on the GeMn granular thin films and decomposed the XMCD signal into two components: the FM component, which can be attributed to the Mn atoms in the Mn-rich nanoparticles, and the PM component, which originates from the Mn atoms in the Mn-poor matrix. We revealed that the MR ratio is proportional to the product of the FM and PM components of the XMCD intensity when $T < T_p$, percolation temperature. Below $T_p$, the spin polarized holes in the FM nanoparticles penetrate into the Mn-poor matrix. Thus, the large MR can be associated with the spin disorder magnetic scattering of these extended spin-polarized holes by the PM Mn atoms in the Mn-poor matrix. Because the MR ratio of the $Ge_{0.86}Mn_{0.14}$ film is about ten times as large as that reported in (In,Mn)Sb, the spin polarization of the holes in the FM nanoparticles is expected to be significantly larger than that (10-30%) in (In,Mn)Sb. The large spin polarization makes $Ge_{1-x}Mn_x$ a promising material for future spintronic applications.

## ACKNOWLEDGEMENTS

This work was partly supported by Grants-in-Aid for Scientific Research (22224005, 23000010, 26249039, and 15H02109) including the Specially Promoted Research, Project for Developing Innovation Systems from MEXT, the Cooperative Research Project Program of RIEC, Tohoku University, and Spintronics Research Network of Japan. This work was performed under the Shared Use Program of JAEA Facilities (Proposal No. 2015B-E24) with the approval of the Nanotechnology Platform Project supported by MEXT. The synchrotron radiation experiments were performed at the JAEA beamline BL23SU in SPring-8 (Proposal No. 2015B3881). Y. K. W. acknowledges financial support from JSPS through the Program for Leading Graduate Schools (MERIT) and the JSPS Research Fellowship Program for Young Scientist. S. S. acknowledges financial support from JSPS through the Program for Leading Graduate Schools (ALPS).

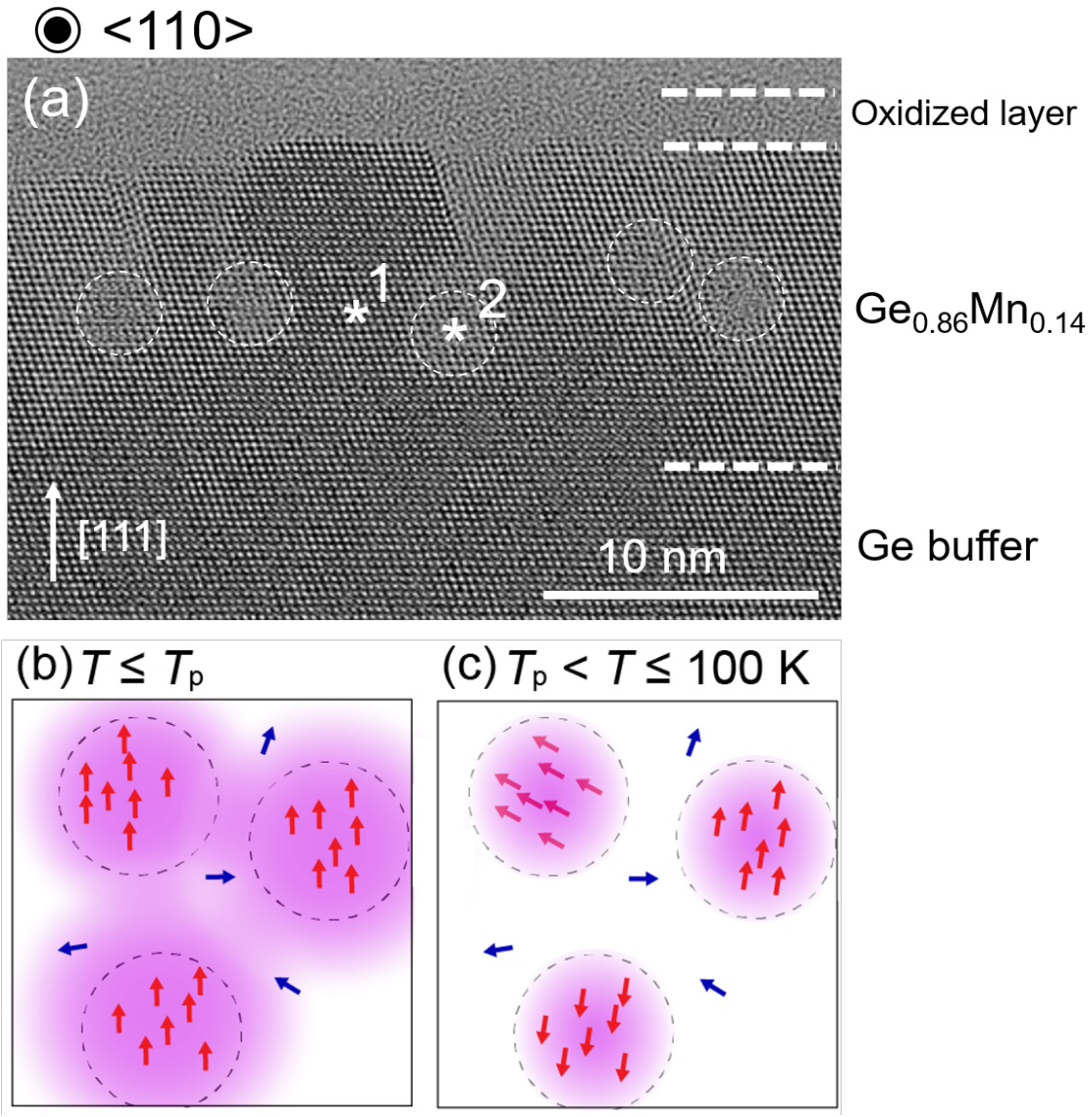

Fig. 1. (a) Transmission-electron-microscope lattice image of the $Ge_{0.86}Mn_{0.14}$ layer projected along the Ge<110> axis. The nanoparticles are indicated by white dashed circles. By the spatially resolved energy dispersive X-ray spectroscopy, the local Mn concentrations at *1 (matrix) and *2 (nanoparticle) are roughly estimated to be ~6% and ~60%, respectively. (b)(c) Schematic illustration of the spatial distribution of the spin-polarized holes (pink regions) originating from the Mn-rich nanoparticles (black dashed circles) when temperature $T \leq T_p$ (b) and $T_p < T \leq 100$ K (c). The red and blue arrows correspond to the magnetic moments of the Mn atoms in the FM nanoparticles and PM matrix, respectively.

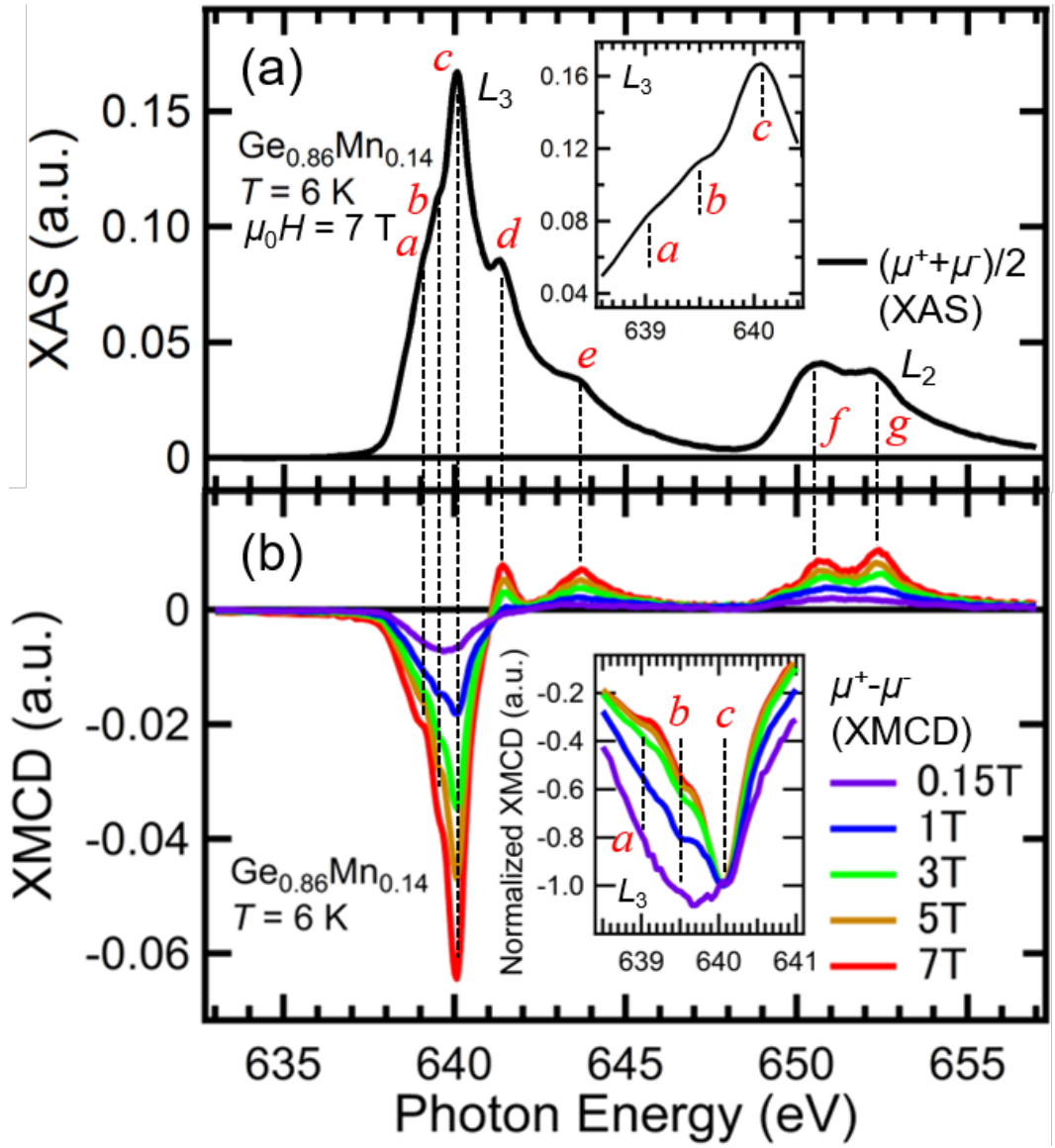

Fig. 2. (a) Mn $L_{2,3}$-edge XAS [$(\mu^+ + \mu^-)/2$] spectrum for the $Ge_{0.86}Mn_{0.14}$ film at 6 K with a magnetic field $\mu_0 H = 7$ T applied perpendicular to the film surface. The inset shows a magnified plot of the spectra at the Mn $L_3$ edge. (b) Mn $L_{2,3}$-edge XMCD (= $\mu^+ - \mu^-$) spectra for the $Ge_{0.86}Mn_{0.14}$ film at 6 K with various magnetic fields $H$ applied perpendicular to the film surface. The inset shows a magnified plot of the spectra at the Mn $L_3$ edge. Here, the XMCD data have been normalized at $c$.

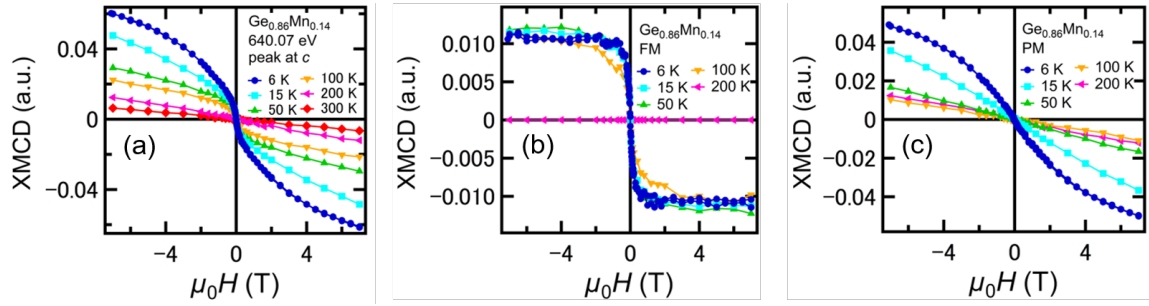

Fig. 3. (a)-(c) Experimentally obtained XMCD-$H$ curves (a) and derived FM (b) and PM (c) components of the XMCD-$H$ curves for the $Ge_{0.86}Mn_{0.14}$ film at various temperatures.

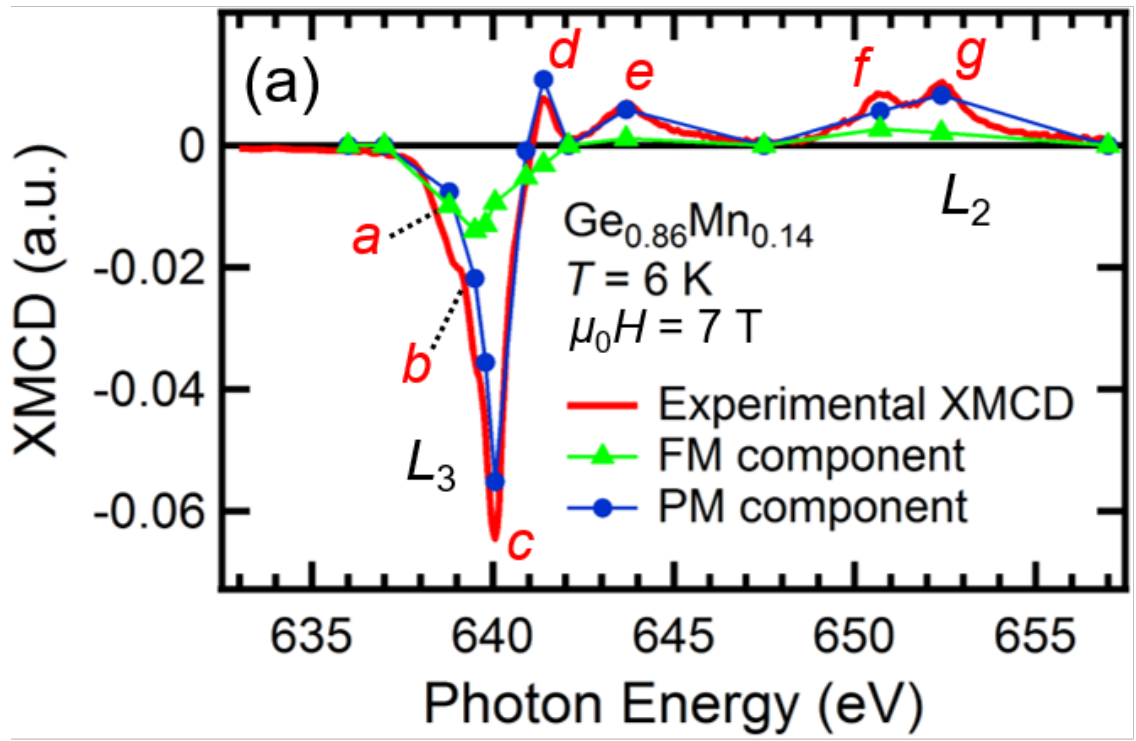

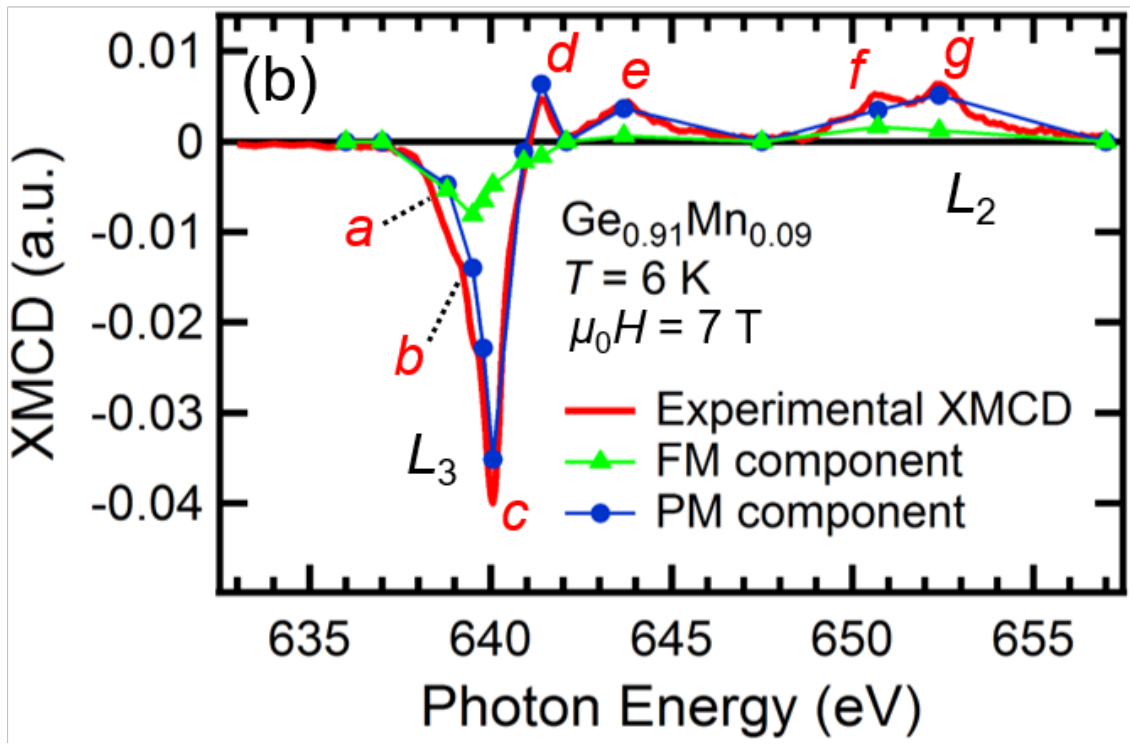

Fig. 4. (Color Online) (a),(b) Experimentally obtained XMCD spectra (red curve), and derived FM (green triangles) and PM (blue circles) components of the XMCD spectra at 6 K with $\mu_0 H$ = 7 T applied perpendicular to the film surface for the $Ge_{0.86}Mn_{0.14}$ film (a) and $Ge_{0.91}Mn_{0.09}$ film (b).

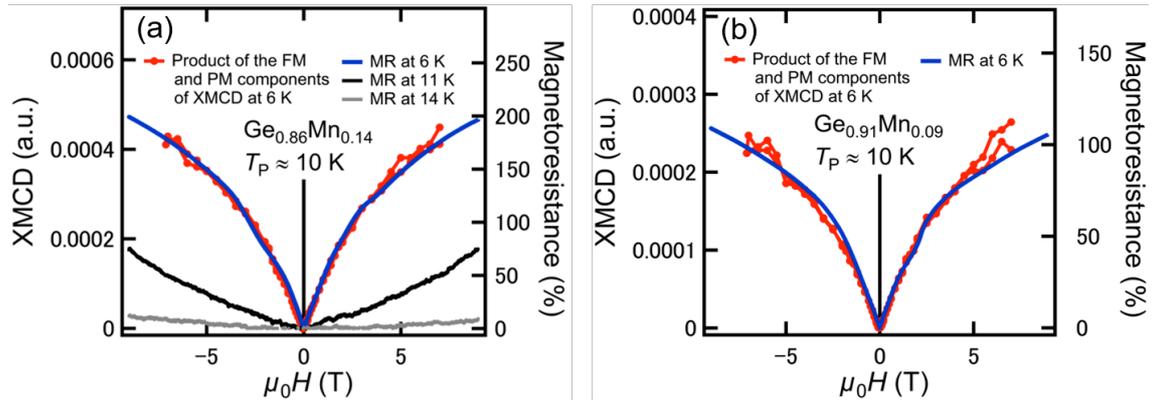

Fig. 5. (a),(b) MR ratio (solid curves) as a function of $\mu_0 H$ applied perpendicular to the film surface for the $Ge_{0.86}Mn_{0.14}$ film (a) and $Ge_{0.91}Mn_{0.09}$ film (b). The magnetic-field dependence of the product of the FM and PM components of the XMCD intensity is also plotted (red points and curves).

**Supplemental Material for Origin of the large positive magnetoresistance**
**in Ge₁₋ₓMnₓ granular thin films**

Yuki K. Wakabayashi,[1] Ryota Akiyama,[2] Yukiharu Takeda,[3] Masafumi Horio,[2] Goro Shibata,[2] Shoya Sakamoto,[2] Yoshisuke Ban,[1] Yuji Saitoh,[3] Hiroshi Yamagami,[3,4] Atsushi Fujimori,[2] Masaaki Tanaka,[1,5] and Shinobu Ohya[1,5]

[1]*Department of Electrical Engineering and Information Systems, The University of Tokyo, 7-3-1 Hongo, Bunkyo-ku, Tokyo 113-8656, Japan*
[2]*Department of Physics, The University of Tokyo, Bunkyo-ku, Tokyo 113-0033, Japan*
[3]*Materials Sciences Research Center, JAEA, Sayo, Hyogo 679-5148, Japan*
[4]*Department of Physics, Kyoto Sangyo University, Motoyama, Kamigamo, Kita-Ku, Kyoto 603-8555, Japan*
[5]*Center for Spintronics Research Network, Graduate School of Engineering, The University of Tokyo, 7-3-1 Hongo, Bunkyo-ku, Tokyo 113-8656, Japan*

## I. Sample preparation

The Ge₁₋ₓMnₓ thin films with the total Mn concentration $x$ of 0.09 and 0.14 were grown on Ge(111) substrates by low-temperature molecular beam epitaxy (LT-MBE). Figures S1(a) and S1(b) show the schematic cross-sectional structures of the samples used for the measurements of the X-ray magnetic circular dichroism (XMCD), superconducting quantum interference device (SQUID) and high-resolution transmission-electron microscopy (HRTEM), and those used for the magneto-transport measurements, respectively. The growth process is described as follows. After the $p$ (or $n$)-type Ge(111) substrates were chemically cleaned by ultra-pure water and acetone, followed by etching with ultra-pure water and buffered HF in a cyclical manner for 1 hour, they were introduced in our ultrahigh vacuum ($\leq \sim 7.0 \times 10^{-9}$ Pa) MBE growth chamber through an oil-free load-lock system. After degassing the substrate at 400°C for 30 minutes and successive thermal cleaning at 740°C for 15 minutes, we grew a 10-nm-thick Ge buffer layer at 180°C, which was followed by the growth of a 13-nm-thick Ge₁₋ₓMnₓ layer at 130°C. After that, in the case of the samples used for the XMCD, SQUID, and HRTEM measurements, we grew a 1.5-nm-thick Ge capping layer at 130°C to avoid the surface oxidation of the GeMn layer. During the growth, the *in situ* reflection high-energy electron diffraction of the Ge₀.₉₁Mn₀.₀₉ and Ge₀.₈₆Mn₀.₁₄ layers showed 2×2 and 1×1 streaks, respectively. This result indicates that the Ge₁₋ₓMnₓ layers are epitaxially grown on the Ge substrates.

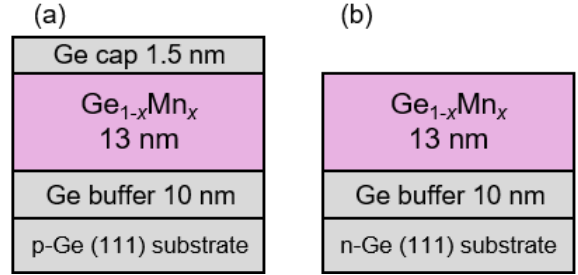

Fig. S1. (a),(b) Schematic cross-sectional structures of the samples used for the XMCD, SQUID and HRTEM measurements (a), and for the magneto-transport measurements (b).

## II. Crystallographic analyses

The crystallographic analyses on the $Ge_{0.86}Mn_{0.14}$ film were performed by HRTEM combined with the spatially resolved transmission electron diffraction (TED) and energy-dispersive X-ray spectroscopy (EDX). From the HRTEM lattice image of the $Ge_{0.86}Mn_{0.14}$ [see Fig. 1(a) in the main text], it is found that the $Ge_{0.86}Mn_{0.14}$ layer basically has a diamond-type crystal structure. There are sphere-like nanoparticles embedded in the matrix [see the dashed circles in Fig. 1(a) in the main text]. By the spatially resolved EDX measurements, the local Mn concentrations at *1 (matrix) and *2 (nanoparticle) are roughly estimated to be ~6% and ~60%, respectively. Figures S2(a) and S2(b) show the TED images at *1 and *2, respectively. The main diffraction patterns at both points indicate the diamond-type crystal structure. Additionally, a weak halo pattern is seen at *2, indicating the presence of amorphous structures. Such a weak halo pattern is also seen in the plane-view TED patterns as shown in Ref. S1. The XRD diffraction patterns of our $Ge_{0.86}Mn_{0.14}$ film indicate the presence of $Mn_5Ge_3$ precipitates [S1], which are the most stable compound in the Mn-Ge phase diagram [S2,S3]. These results indicate that the Mn-rich nanoparticles are a heavily-Mn doped amorphous GeMn phase including $Mn_5Ge_3$ precipitates, and that the surrounding Mn-poor matrix has a diamond-type crystal structure.

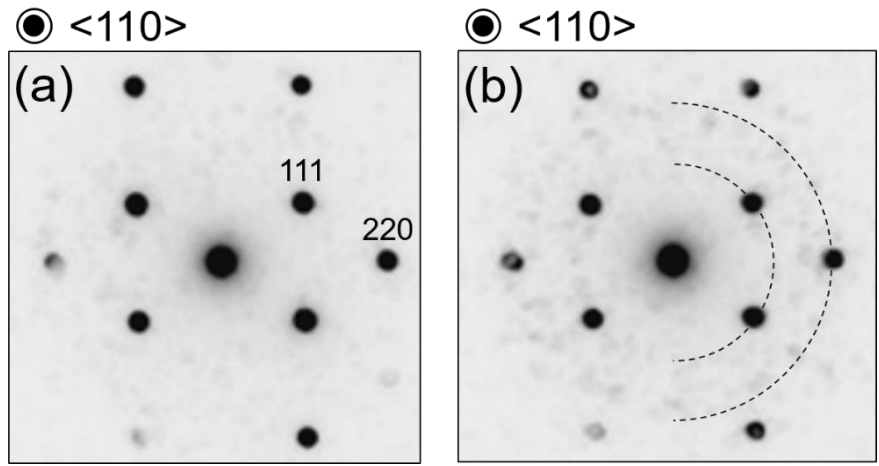

Fig. S2. (a), (b) TED images at *1 (matrix) (a) and *2 (nanoparticle) (b) projected along the Ge<110> axis. The weak halo pattern is indicated by the dashed curves in (b).

## III.    X-ray absorption spectroscopy (XAS) and XMCD spectra of the Ge$_{0.91}$Mn$_{0.09}$ film

Figure S3(a) shows the Mn $L_{2,3}$-edge XAS [$(\mu^+ + \mu^-)/2$] spectrum obtained at 6 K with a magnetic field $\mu_0 H = 7$ T applied perpendicular to the film surface for the Ge$_{0.91}$Mn$_{0.09}$ film. The spectrum has five peaks, whose energies are referred to as $a$-$e$, at the Mn $L_3$-edge [see also the inset in Fig. S3(a)] and two peaks, whose energies are referred to as $f$ and $g$, at the Mn $L_2$-edge. These energy positions are the same as those observed for the Ge$_{0.86}$Mn$_{0.14}$ film [see Fig. 2(a) in the main text]. Figure S3(b) shows the Mn $L_{2,3}$-edge XMCD ($= \mu^+ - \mu^-$) spectra for the Ge$_{0.91}$Mn$_{0.09}$ film at 6 K with various magnetic fields $H$ applied perpendicular to the film surface. We can see clear peaks at $a$-$g$. When normalized to 640.06 eV (peak at $c$), the XMCD spectra with various $H$ differ, and the peak at $c$ becomes more dominant as $H$ increases, as shown in the inset of Fig. S3(b).

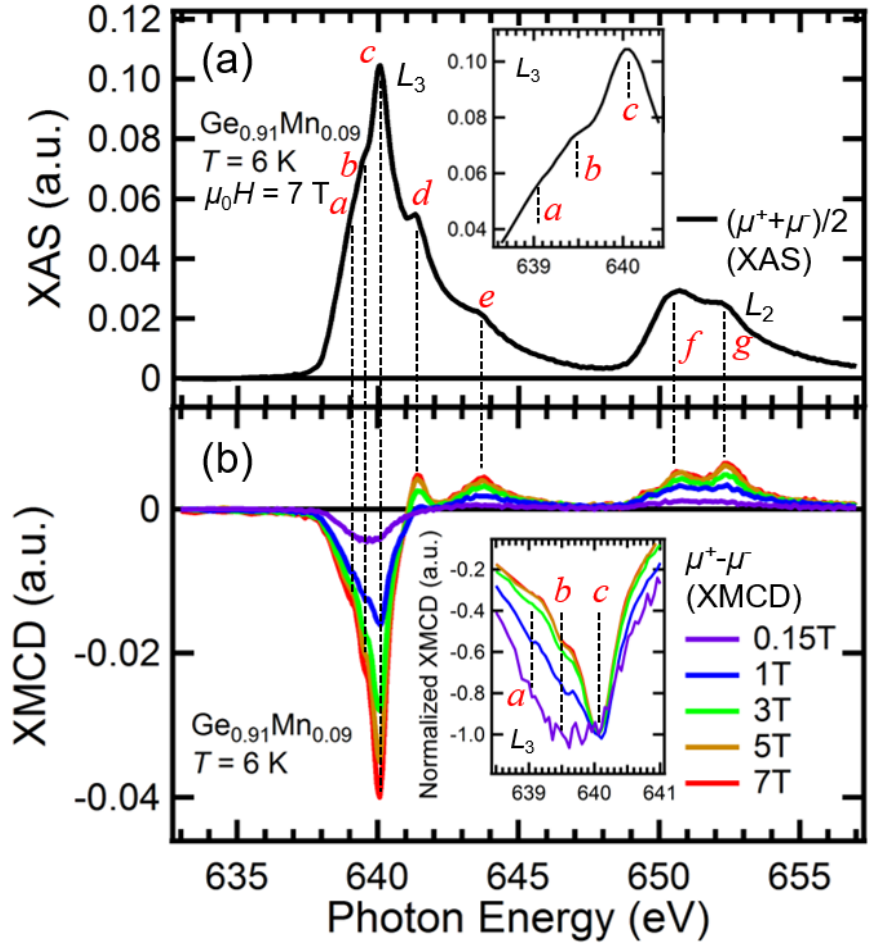

Fig. S3. (a) Mn $L_{2,3}$-edge XAS [$(\mu^+ + \mu^-)/2$] spectrum for the Ge$_{0.91}$Mn$_{0.09}$ film at 6 K with $\mu_0 H = 7$ T applied perpendicular to the film surface. The inset shows a magnified plot of the spectrum at the Mn $L_3$ edge. (b) Mn $L_{2,3}$-edge XMCD ($= \mu^+ - \mu^-$) spectra for the Ge$_{0.91}$Mn$_{0.09}$ film at 6 K with various $H$ applied perpendicular to the film surface. The inset shows a magnified plot of the spectra at the Mn $L_3$ edge, where the XMCD data are normalized to 640.06 eV. The black dashed lines indicate the peak positions.

## IV. Derivation of the ferromagnetic (FM) and paramagnetic (PM) components of the XMCD-$H$ curves

Figures S4(a) and S4(b) show the $H$ dependence of the XMCD intensity measured at 6 K for the $Ge_{0.86}Mn_{0.14}$ film at $a$ and $c$, respectively. These curve shapes are largely different. These curves are composed of an FM-like component $I_{FM}(H)$, which saturates at high magnetic fields ($> 6$ T), and a PM component $I_{PM}(H)$, which is linear in the range of $\mu_0H$ from $-1$ T to 1 T and follows the Langevin function. That is,

$$I_{XMCD}(H, a) = \alpha(a) \ I_{FM}(H) + \beta(a) \ I_{PM}(H) \tag{S1}$$

$$I_{XMCD}(H, c) = \alpha(c) \ I_{FM}(H) + \beta(c) \ I_{PM}(H), \tag{S2}$$

where $I_{XMCD}(H, E)$ is the XMCD intensity at an energy $E$ under $H$, and $\alpha(E)$ and $\beta(E)$ are $E$ dependent constants. Here, $I_{FM}(H)$ and $I_{PM}(H)$ are normalized to 1 at 7 T.

Here, we determine $I_{FM}(H)$ and $I_{PM}(H)$ from the XMCD-$H$ curves at $a$ and $c$. By subtracting Eq.(S2)$\times \beta(a)/\beta(c)$ from Eq.(S1), we obtain

$$I_{XMCD}(H, a) - \gamma I_{XMCD}(H, c) = \lambda I_{FM}(H). \tag{S3}$$

Here,

$$\gamma = \frac{\beta(a)}{\beta(c)}, \qquad \lambda = \alpha(a) - \alpha(c)\frac{\beta(a)}{\beta(c)}.$$

$\lambda$ is just a constant for normalization of $I_{FM}(H)$. We determined $\gamma$ so that $I_{XMCD}(H, a) - \gamma I_{XMCD}(H, a)$, i.e. $I_{FM}(H)$, saturates when $\mu_0H > 6$ T. Then, we can derive $I_{FM}(H)$ (Fig. S4(c)).

Similarly, for the determination of $I_{PM}(H)$, we obtain

$$I_{XMCD}(H, c) - \eta I_{XMCD}(H, a) = \xi I_{PM}(H) \tag{S4}$$

from Eqs. (S1) and (S2). Here,

$$\eta = \frac{\alpha(c)}{\alpha(a)}, \qquad \xi = \beta(c) - \beta(a)\frac{\alpha(c)}{\alpha(a)}.$$

We determined $\xi$ so that $I_{XMCD}(H, a) - \eta I_{XMCD}(H, a)$, i.e. $I_{PM}(H)$, becomes linear in the range of $\mu_0H$ from $-1$ T to 1 T. This is because the magnetization curve of paramagnetic spins is generally linear in this magnetic field range (from $-1$ to 1 T) as long as the moment of the spin is lower than $10\mu_B$. Then, we can derive $I_{PM}(H)$ (Fig. S4(d)).

The FM component is attributed to the FM Mn atoms in the Mn-rich nanoparticles. The derived PM component of the XMCD-$H$ curve follows the Langevin function with the magnetic moment of 5 $\mu_B$, as shown in Fig. S4(d). Thus, the PM component of the XMCD signal originates from the PM $Mn^{2+}$ ions with $n_{3d} = 5$ (5 $\mu_B$), where $n_{3d}$ is the number of 3d electrons per $Mn^{2+}$.

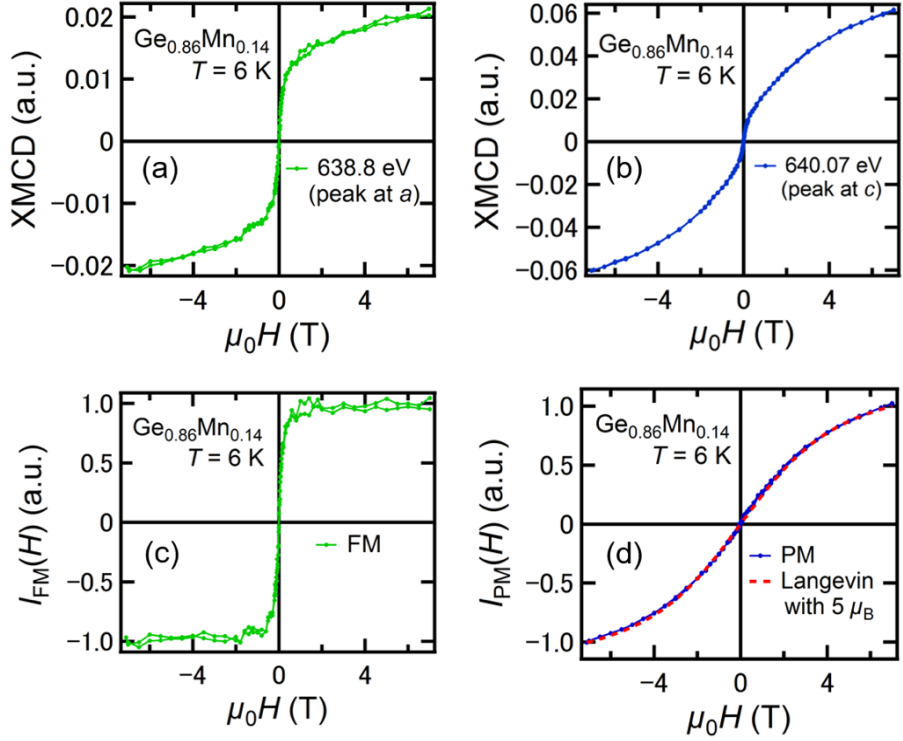

Fig. S4. (a),(b) $H$ dependence of the XMCD intensity measured at 6 K for the $Ge_{0.86}Mn_{0.14}$ film at 638.8 eV (peak at $a$) (a) and 640.07 eV (peak at $c$) (b). (c),(d) Derived FM (c) and PM (d) components of the XMCD-$H$ curves for the $Ge_{0.86}Mn_{0.14}$ film. The red dashed curve is the Langevin function with the magnetic moment of 5 $\mu_B$ at 6 K.

Figures S5(a)-S5(g) show the XMCD-$H$ curves (red solid curves) measured at 6 K for the $Ge_{0.86}Mn_{0.14}$ film at various energies. These XMCD-$H$ curves show various shapes depending on $E$. As shown in Figs. S5(a)-S5(g), the experimental XMCD-$H$ curves at various energies are well fitted by the linear combination of $I_{FM}(H)$ and $I_{PM}(H)$, which were derived above (black dashed curves). Using the same procedure, the XMCD-$H$ curves at various energies also for the $Ge_{0.91}Mn_{0.09}$ film are well fitted by the linear combination of $I_{FM}(H)$ and $I_{PM}(H)$ derived using the XMCD-$H$ curves at $a$ and $c$ for the $Ge_{0.91}Mn_{0.09}$ film (see Figs. S6 and S7). These results indicate that the XMCD spectra are entirely composed only of the FM-like and PM components derived above.

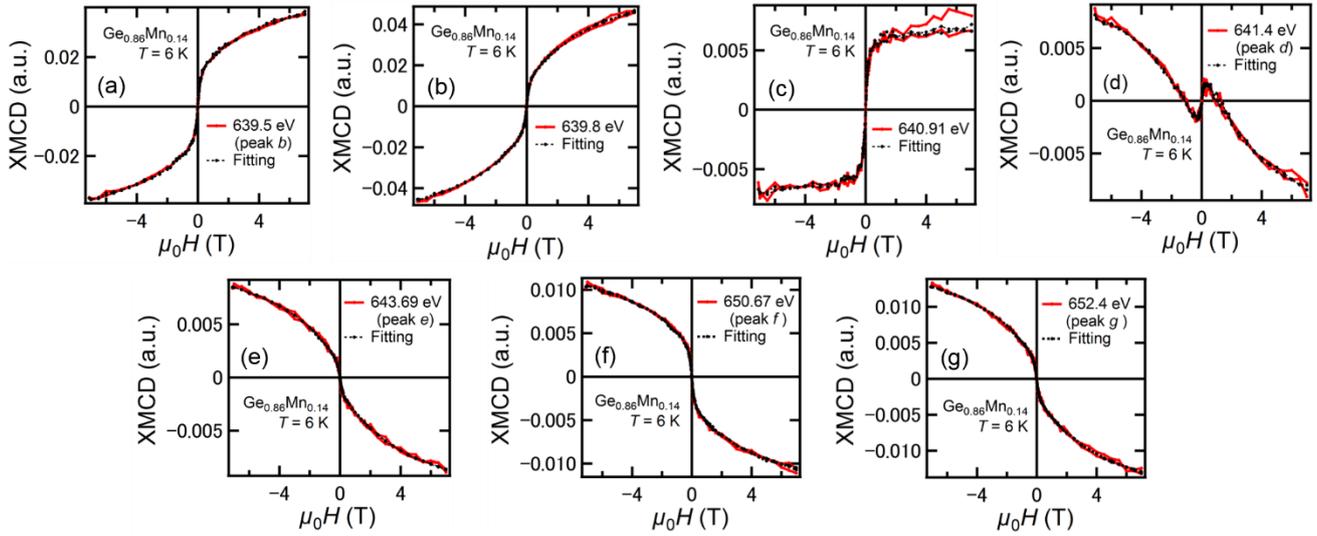

Fig. S5. Experimental XMCD-*H* curves (red solid curves) at various energies at 6 K and the fitting curves (black dashed curves) expressed by the linear combination of $I_{FM}(H)$ and $I_{PM}(H)$ for the $Ge_{0.86}Mn_{0.14}$ film.

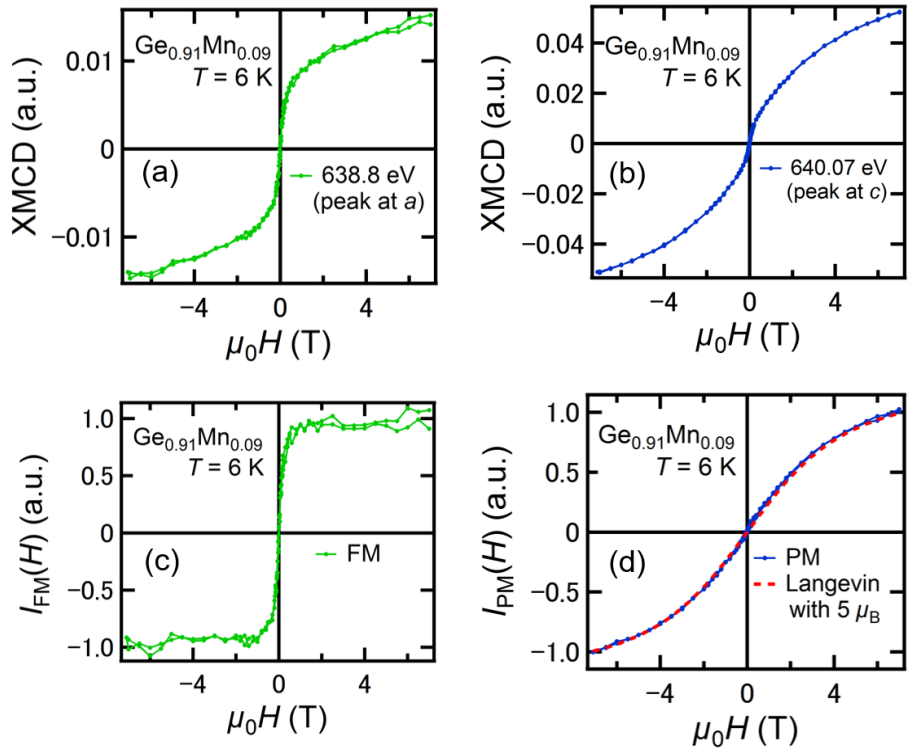

Fig. S6. (a),(b) *H* dependence of the XMCD intensities measured at 6 K for the $Ge_{0.91}Mn_{0.09}$ film at 638.8 eV (peak at *a*) (a) and 640.07 eV (peak at *c*) (b). (c),(d) Derived FM (c) and PM (d) components of the XMCD-*H* curves for the $Ge_{0.91}Mn_{0.09}$ film. The red dashed curve is the Langevin function with the magnetic moment of 5 $\mu_B$ at 6 K.

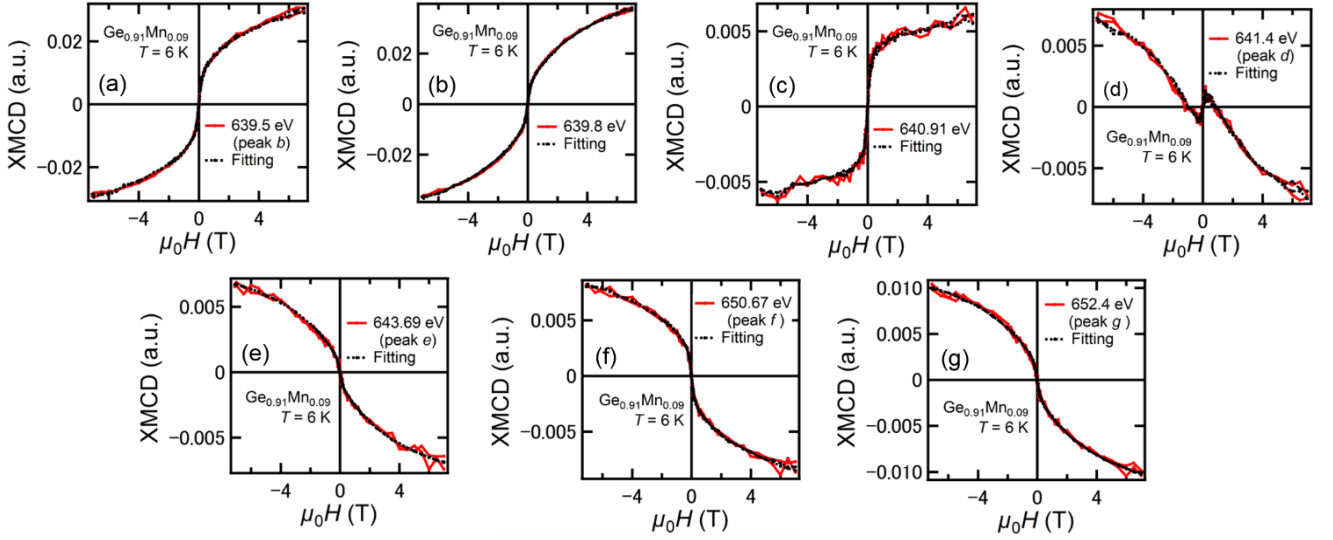

Fig. S7. Experimental XMCD-*H* curves (red solid curves) at various energies at 6 K and the fitting curves (black dashed curves) expressed by the linear combination of $I_{\mathrm{FM}}(H)$ and $I_{\mathrm{PM}}(H)$ for the $Ge_{0.91}Mn_{0.09}$ film.

## V. Experimentally obtained XMCD-*H* curves and derived FM and PM components of the XMCD-*H* curves for the $Ge_{0.91}Mn_{0.09}$ film at various temperatures

Figures S8(a)-S8(c) show the experimentally obtained XMCD-*H* curves (a) and derived FM (b) and PM components (c) of the XMCD-*H* curves for the $Ge_{0.91}Mn_{0.09}$ film at various temperatures.

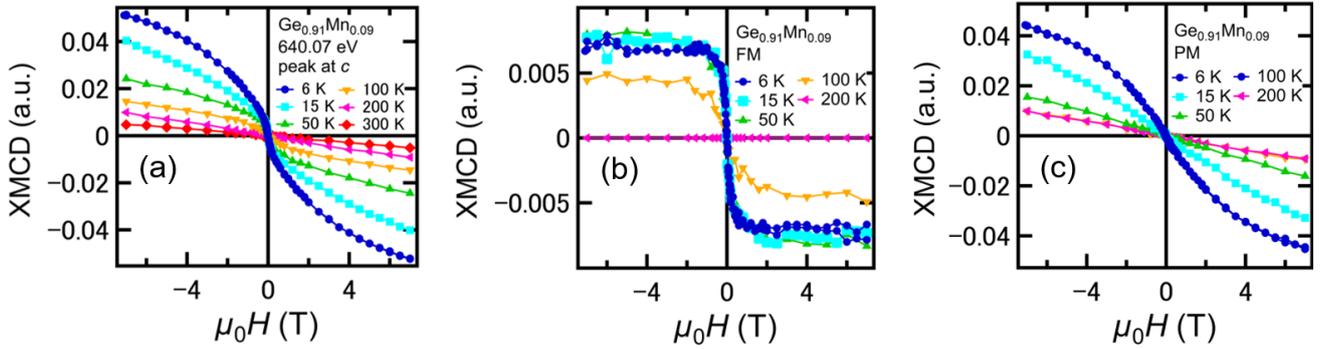

Fig. S8. (a)-(c) Experimentally obtained XMCD-*H* curves (a), and derived FM (b) and PM components (c) of the XMCD-*H* curves for the $Ge_{0.91}Mn_{0.09}$ film at various temperatures.

## VI. Curie plots of the PM component of the XMCD intensity

For both $Ge_{0.86}Mn_{0.14}$ and $Ge_{0.91}Mn_{0.09}$ films, the Curie plots of the PM component of the XMCD intensity are linear as a function of the temperature (Fig. S9). This is typical paramagnetic behavior, which confirms that our method of the decomposition of the XMCD intensity into the FM and PM components is appropriate.

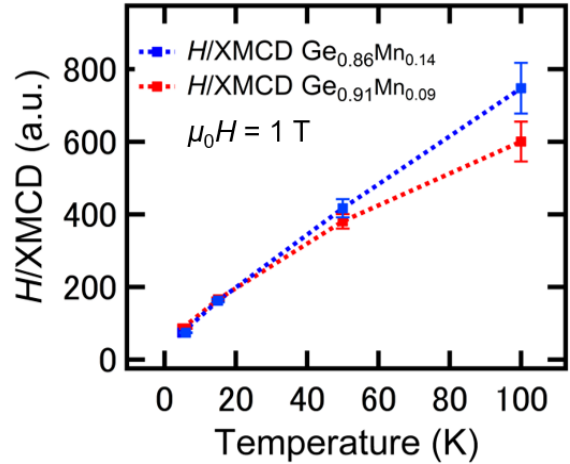

Fig. S9. Temperature dependence of $H$/XMCD of the PM matrix with $\mu_0 H = 1$ T applied perpendicular to the film surface for the $Ge_{0.86}Mn_{0.14}$ film (blue symbols) and $Ge_{0.91}Mn_{0.09}$ film (red symbols).

## VII. Derivation of the XAS spectra of the FM nanoparticles and of the PM matrix

In order to derive the XAS spectra of the FM nanoparticles and PM matrix, we fitted the sum of the Gaussian peaks located at $a$-$g$ to the XAS spectra measured for the $Ge_{0.86}Mn_{0.14}$ film at 6 K with $\mu_0 H = 7$ T applied perpendicular to the film surface [Fig. S10(a)]. Here, we decompose the fitting spectrum into that of the FM nanoparticles and that of the PM matrix. At the Mn $L_3$ edge, the Gaussian peaks at $a$, $b$ are attributed to the FM nanoparticles because the FM component of the XMCD intensity is strongest at $b$ and because the FM component of the XMCD intensity is stronger than the PM component at $a$ [see Fig. 3(a) in the main text]. At the Mn $L_3$ edge, the Gaussian peaks at $c$, $d$, and $e$ are attributed to the PM matrix because the PM component of the XMCD intensity is stronger than the FM component at $c$, $d$, and $e$ (see Fig. 4 in the main text). At the Mn $L_2$ edge, the Gaussian peak at $f$ ($g$) is attributed to the FM nanoparticles (the PM matrix) because the FM component (PM component) of the XMCD intensity has a peak at $f$ ($g$) (see Fig. 4 in the main text). Thus, the XAS spectrum of the FM nanoparticles (PM matrix) is expressed as a sum of the Gaussian peaks at $a$, $b$, and $f$ ($c$, $d$, $e$, and $g$). Figure S10(b) shows the fitting XAS spectrum [this is the same curve as that shown in Fig. S10(a)], and the derived XAS spectra of the FM nanoparticles and PM matrix for the $Ge_{0.86}Mn_{0.14}$ film. Similarly to the derived FM (PM) component of the XMCD spectra (see Fig. 4 in the main text), the derived XAS spectrum of the FM nanoparticles (PM matrix) has a broad peak composed of the Gaussian peaks at $a$ and $b$ (multiple peaks at $c$, $d$, and $e$) at the Mn $L_3$-edge. This result indicates that the $3d$ electrons of the FM Mn atoms are *not* localized at each Mn atom and that those of the PM Mn atoms have a localized nature. We also decomposed the XAS spectra into those of the FM nanoparticles and PM matrix for the $Ge_{0.91}Mn_{0.09}$ film using the same procedure. The same features mentioned above were observed for the $Ge_{0.91}Mn_{0.09}$ film as shown in Fig. S11.

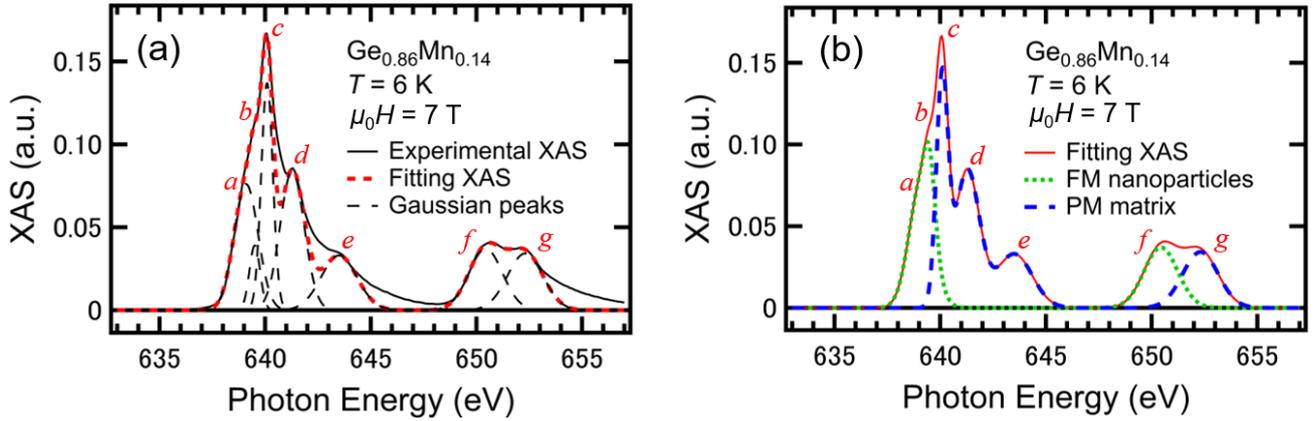

Fig. S10. (a) Experimental XAS spectrum (black curve) measured at 6 K with $\mu_0H = 7$ T applied perpendicular to the film surface for the $Ge_{0.86}Mn_{0.14}$ film, and fitting XAS spectrum (red dashed curve) expressed by the sum of the Gaussian peaks located at *a-g* (black dashed curves). (b) Fitting XAS spectrum (red curve) and derived XAS spectra of the FM nanoparticles (green dashed curve) and PM matrix (blue dashed curve).

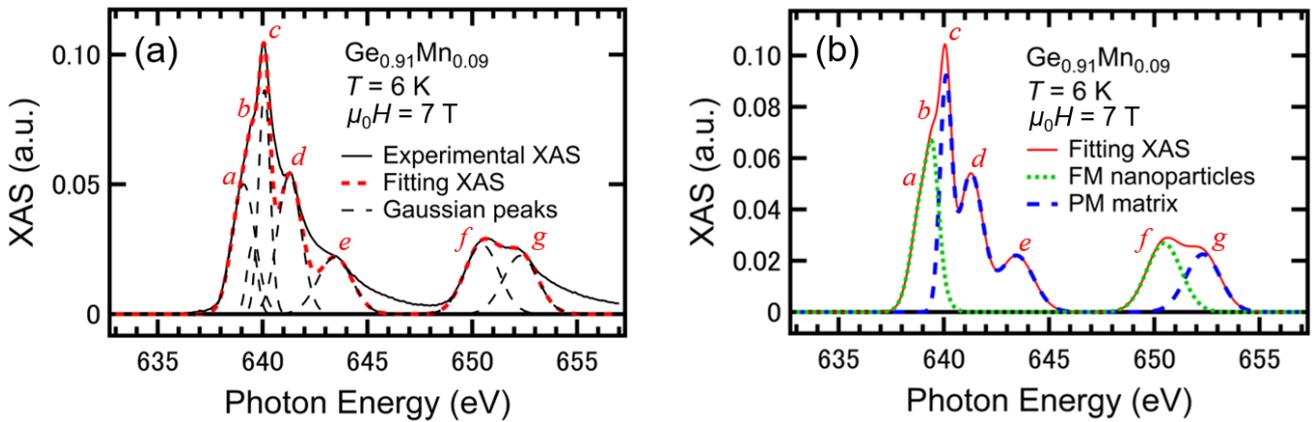

Fig. S11. (a) Experimental XAS spectrum (black curve) measured at 6 K with $\mu_0H = 7$ T applied perpendicular to the film surface for the $Ge_{0.91}Mn_{0.09}$ film, and fitting XAS spectrum (red dashed curve) expressed by the sum of the Gaussian peaks located at *a-g* (black dashed curves). (b) Fitting XAS spectrum (red curve) and derived XAS spectra of the FM nanoparticles (green dashed curve) and PM matrix (blue dashed curve).

## VIII. Estimation of the orbital magnetic moment, $m_{orb}$, and the spin magnetic moment, $m_{spin}$, of the FM Mn atoms in the Mn-rich nanoparticles and the PM $Mn^{2+}$ ions in the Mn-poor matrix

We can see characteristic features of the Mn-rich nanoparticles and Mn-poor matrix from the orbital magnetic moment, $m_{orb}$, and the spin magnetic moment, $m_{spin}$, of each region (Table S1), which are obtained by the well-established procedure using the XMCD sum rules separately [S4-S6]. Figure S12(a) shows the derived XAS spectrum of the FM nanoparticles (solid curve) and its integration from 635 eV (dashed curve) for the $Ge_{0.86}Mn_{0.14}$ film. Figure S12(b) shows the derived FM component of the XMCD spectrum (solid curve) and its integration from 635 eV (dashed curve) for the $Ge_{0.86}Mn_{0.14}$ film. Figures S13(a) and S13(b) show the same data for the PM matrix. For the XMCD sum-rules analyses, we define $r$, $p$, and $q$ as the following equations.

$$r = \int_{E_3+E_2} \frac{(\mu^+ + \mu^-)}{2} dE, \quad (S5)$$

$$p = \int_{E_3} (\mu^+ - \mu^-) dE, \quad (S6)$$

$$q = \int_{E_3+E_2} (\mu^+ - \mu^-) dE, \quad (S7)$$

where $E_3$ (635-648 eV) and $E_2$ (648-665 eV) represent the integration energy ranges for the $L_3$ and $L_2$ absorption edges, respectively. We used the XMCD sum rules, which are expressed as follows:

$$m_{orb} = -\frac{2q}{3r}(10 - n_{3d}), \quad (S8)$$

$$m_{spin} + 7m_T = -\frac{3p-2q}{r}(10 - n_{3d}), \quad (S9)$$

where $n_{3d}$ and $m_T$ are the number of $3d$ electrons of the Mn atom and the expectation value of the intra-atomic magnetic dipole operator, respectively. For the $Mn^{2+}$ ions in the Mn-poor matrix, we took $n_{3d}$ to be 5 and the correction factor for $m_{spin}$ to be 0.68 [S7]. We neglected $m_T$ for the PM $Mn^{2+}$ ions in the Mn-poor matrix because it is negligibly small at the $T_d$ symmetry site [S8]. For the Mn atoms in the FM nanoparticles, because the valence is unknown, we took $n_{3d}$ to be $4-6$ and the correction factor for $m_{spin}$ to be from –0.5 to 0.5 [S9]. We neglected $m_T$ for the Mn atoms in the FM Mn-rich nanoparticles because these regions have sphere-like shapes [see Fig. 1(a) in the main text] [S10]. As seen in Table S1, for both samples, the $m_{spin}$ value of the PM $Mn^{2+}$ ions (~2.4 $\mu_B$) is lower than the ideal $m_{spin}$ value of $Mn^{2+}$ (i.e. 5 $\mu_B$). This suggests that some of the Mn atoms in the Mn-poor matrix are magnetically inactive [S11-S13]. The large $m_{orb}/m_{spin}$ (= $0.12-0.39$) value of the FM Mn atoms is a characteristic property observed in magnetic nanoparticles [S14]. By contrast, the $m_{orb}$ value of the PM matrix vanishes for both samples, confirming that the valence of the PM Mn atoms is 2+ with $n_{3d}$ = 5. The $m_{spin}$ and $m_{orb}$ values of the PM $Mn^{2+}$ ions are comparable between $Ge_{0.91}Mn_{0.09}$ and $Ge_{0.86}Mn_{0.14}$. This indicates that the $Mn^{2+}$ ions in the Mn-poor matrix are isolated and that the localized $3d$ state of the $Mn^{2+}$ ions is not affected by the total Mn concentration $x$. On the other hand, for the FM nanoparticles, the $Ge_{0.86}Mn_{0.14}$ film has slightly larger $m_{spin}$ and $m_{orb}$ values than the $Ge_{0.91}Mn_{0.09}$ film. This means that the $3d$ electrons in the FM nanoparticles are delocalized and that they are influenced by the surrounding environment (i.e. local concentration of Mn) because each Mn-rich nanoparticle is locally metallic.

Table S1. The $m_{spin}$, $m_{orb}$, and $m_{orb}/m_{spin}$ values of the Mn atoms in the FM nanoparticles and PM $Mn^{2+}$ ions in the $Ge_{0.91}Mn_{0.09}$ film and $Ge_{0.86}Mn_{0.14}$ film.

| | FM Mn atoms | | PM $Mn^{2+}$ ions | |
| --- | --- | --- | --- | --- |
| | $Ge_{0.91}Mn_{0.09}$ | $Ge_{0.86}Mn_{0.14}$ | $Ge_{0.91}Mn_{0.09}$ | $Ge_{0.86}Mn_{0.14}$ |
| $m_{spin}$ | 0.45-2.05 | 0.56-2.49 | 2.45 | 2.43 |
| $m_{orb}$ | 0.16-0.24 | 0.22-0.32 | 0 | 0 |
| $m_{orb}/m_{spin}$ | 0.12-0.36 | 0.13-0.39 | 0 | 0 |

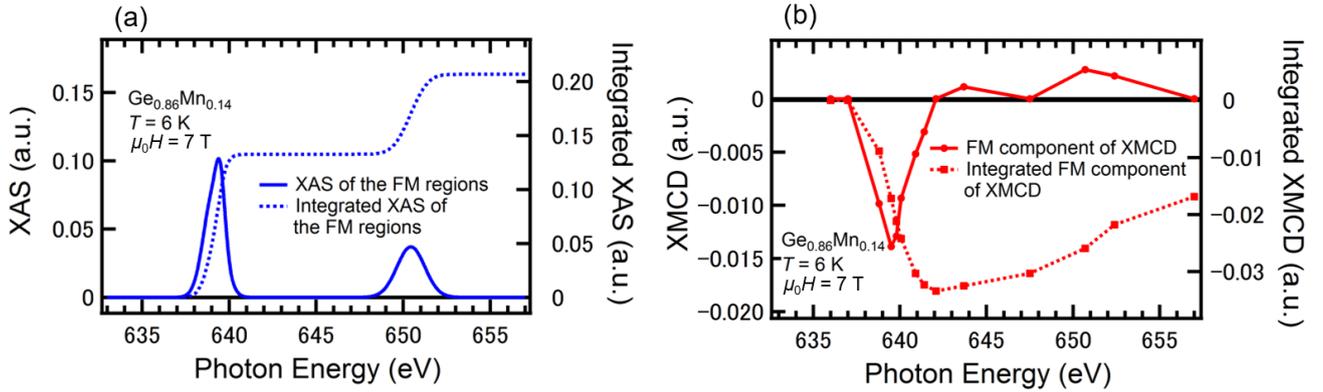

Fig. S12. (a) Derived XAS spectrum of the FM nanoparticles (solid curve) and its integration from 635 eV (dashed curve) for the $Ge_{0.86}Mn_{0.14}$ film. (b) Derived FM component of the XMCD spectrum (solid curve) and its integration from 635 eV (dashed curve) for the $Ge_{0.86}Mn_{0.14}$ film.

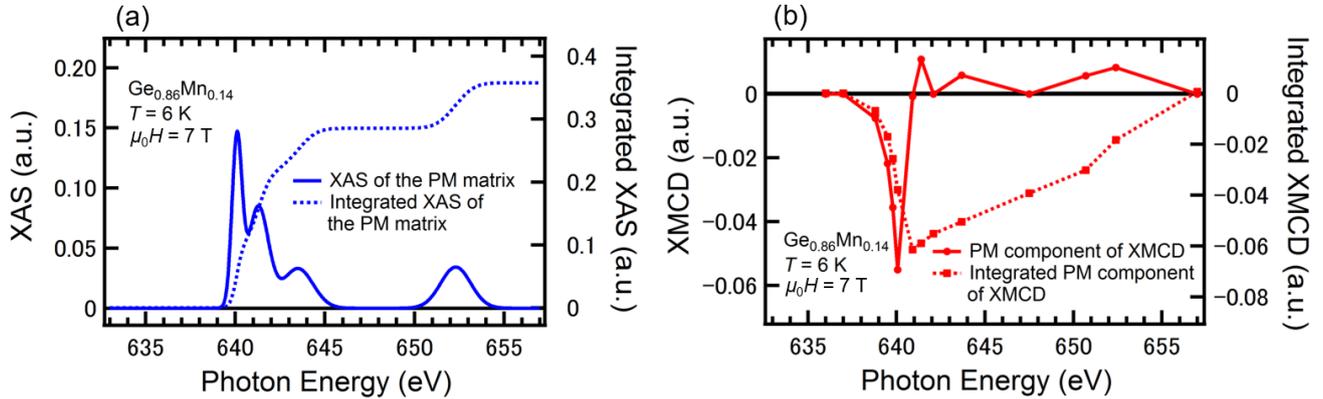

Fig. S13. (a) Derived XAS spectrum of the PM matrix (solid curve) and its integration from 635 eV (dashed curve) for the $Ge_{0.86}Mn_{0.14}$ film. (b) Derived PM component of the XMCD spectrum (solid curve) and its integration from 635 eV (dashed curve) for the $Ge_{0.86}Mn_{0.14}$ film.

## IX. Magnetization curves measured by SQUID

Figures S14(a) and S14(b) show the magnetization curves at various temperatures measured by SQUID for the $Ge_{0.86}Mn_{0.14}$ and $Ge_{0.91}Mn_{0.09}$ films, respectively. The insets show the close-up view near the zero magnetic field. For both samples, clear hysteresis curves were observed below the percolation temperature $T_\mathrm{p} \approx 10$ K. In the $\mathrm{Ge_{0.86}Mn_{0.14}}$ film, a small hysteresis is observed even at 15 K, which is higher than $T_\mathrm{p}$. This probably originates from the spin blocking [S15].

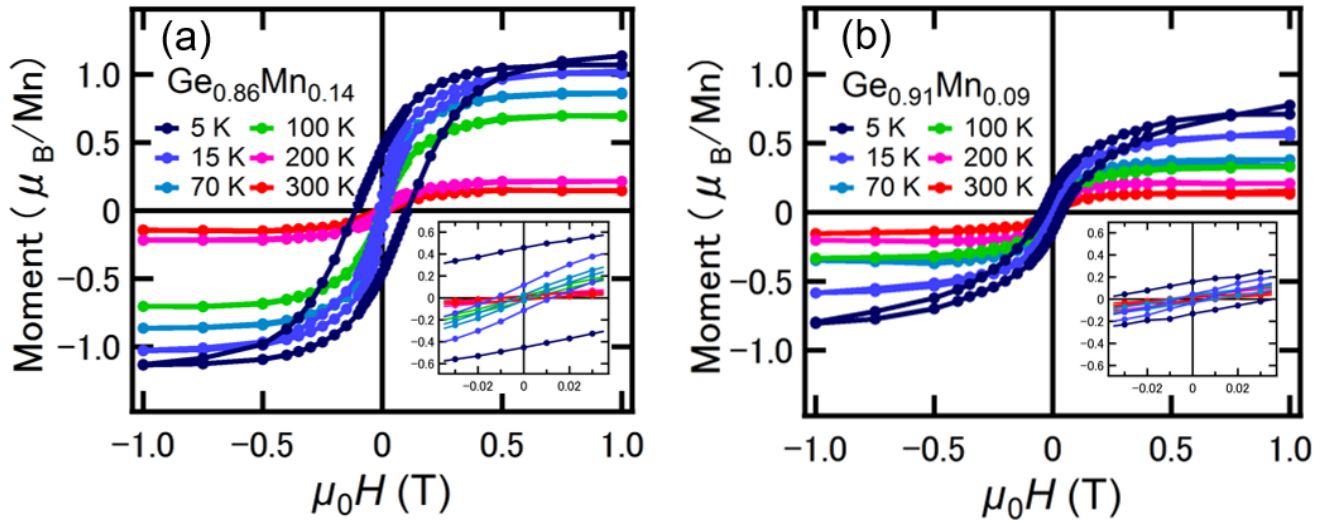

Fig. S14. (a),(b) The $H$ dependence of the magnetization measured by SQUID for the $\mathrm{Ge_{0.86}Mn_{0.14}}$ film (a) and for the $\mathrm{Ge_{0.91}Mn_{0.09}}$ film (b) at various temperatures.